\documentclass[12pt]{iopart}

\usepackage{graphicx}

\usepackage{subfigure}

\usepackage[numbers,sort&compress]{natbib}

\usepackage{indentfirst}

\usepackage{graphicx}
\usepackage{dcolumn}
\usepackage{bm}
\usepackage{epsfig}
\usepackage{subfigure}
\usepackage{graphics}
\usepackage{epstopdf}
\usepackage[english]{babel}
\usepackage{amsfonts}
\usepackage{mathrsfs}
\usepackage{amsthm}
\usepackage{amssymb}
\usepackage{enumerate}
\usepackage{subfigure}
\usepackage{algorithm}
\usepackage{algorithmicx}
\usepackage{algpseudocode}
\usepackage{hyperref}
\usepackage{multirow}
\floatname{algorithm}{Algorithm}

\begin{document}

\title[Zhenyu Shi, Wei Wei, Xiangnan Feng, Xing Li and Zhiming Zheng]{Dynamic aspiration based on Win-Stay-Lose-Learn rule in Spatial Prisoner¡¯s Dilemma Game}

\author{Zhenyu Shi \textsuperscript{1,2,3,4} , Wei Wei \textsuperscript{*,1,2,3,4}, Xiangnan Feng \textsuperscript{*,1,2,3,4}, Xing Li \textsuperscript{1,2,3,4} and Zhiming Zheng \textsuperscript{1,2,3,4}}


\address{1.School of Mathematical Sciences, Beihang University, Beijing, China

2.Key Laboratory of Mathematics Informatics Behavioral Semantics, Ministry of Education, China

3.Peng Cheng Laboratory, Shenzhen, Guangdong, China

4.Beijing Advanced Innovation Center for Big Data and Brain Computing, Beihang

}

\ead{weiw@buaa.edu.cn \textbf{and} fengxiangnan@buaa.edu.cn}
\vspace{10pt}
\begin{indented}
\item[]September 2020
\end{indented}

\begin{abstract}
Prisoner's dilemma game is the most commonly used model of spatial evolutionary game which is considered as a paradigm to portray competition among selfish individuals. In recent years, Win-Stay-Lose-Learn, a strategy updating rule base on aspiration, has been proved to be an effective model to promote cooperation in spatial prisoner's dilemma game, which leads aspiration to receive lots of attention. But in many research the assumption that individual¡¯s aspiration is fixed is inconsistent with recent results from psychology. In this paper, according to Expected Value Theory and Achievement Motivation Theory, we propose a dynamic aspiration model based on Win-Stay-Lose-Learn rule in which individual's aspiration is inspired by its payoff. It is found that dynamic aspiration has a significant impact on the evolution process, and different initial aspirations lead to different results, which are called \emph{Stable Coexistence under Low Aspiration}, \emph{Dependent Coexistence under Moderate aspiration} and \emph{Defection Explosion under High Aspiration} respectively. Furthermore, a deep analysis is performed on the local structures which cause cooperator's existence or defector's expansion, and the evolution process for different parameters including strategy and aspiration. As a result, the intrinsic structures leading to defectors' expansion and cooperators' survival are achieved for different evolution process, which provides a penetrating understanding of the evolution.
Compared to fixed aspiration model,
dynamic aspiration introduces a more satisfactory explanation on  population evolution laws
and can promote deeper comprehension for the principle of prisoner¡¯s dilemma.
\end{abstract}

%
%
%
%
%

\section{Introduction}

The emergence and stability of cooperative behavior among selfish individuals is a challenging problem in biology, social and economic \cite{axelrod1981evolution}. The prisoner¡¯s dilemma game (PDG) is considered as a paradigm to portray competition among selfish individuals \cite{rapoport1965prisoner, axelrod1980prisoner, axelrod1980prisoner2, axelrod1987prisoner, szabo1998prisoner}.
For general parameter settings, defection is better for selfish individuals to survive in population, which is called the \emph{tragedy of the commons} \cite{hardin1968tragedy}. However, we can easily observe numerous cooperation phenomenon in various scenarios. For example, animals will cooperate to obtain food instead of preying alone; companies will set appropriate commodity prices instead of maliciously cutting prices; humans will choose to obey the order instead of jumping in line, etc \cite{Myerson1991Game}. Evolutionary game theory provides an practical framework to explain how the cooperation forms \cite{smith1982evolution, axelrod1984evolution, weibull1995evolution, hofbauer1998evolution, sandholm2006evolution}. Besides, five representative mechanisms considered as promoting cooperative have been investigated: kin selection, direct and indirect reciprocity, network reciprocity and group selection \cite{nowak2006five}.

Since the pioneering work of Nowak and May \cite{nowak1992spatial}, spatial games were proposed and have attracted ample attention of researchers, in which players are located on the spatially structured network and only interact with their neighbors. Since then, numerous studies have emerged to propose various mechanisms which explained the emergence and stability of cooperative behavior, such as punishment \cite{herrmann2008punishment, helbing2010punishment, chen2014punishment, chen2015punishment}, migration \cite{helbing2008migration, cong2012migration, ichinose2013migration}, game organizers \cite{szolnoki2015conformist, szolnoki2016conformist}, teaching ability \cite{szolnoki2007teaching, szolnoki2008teaching, szabo2009teaching}, and so on.
In recent years, aspiration, a value representing individual's expectation, has attracted many researchers¡¯ attention \cite{Yang2012aspiration, Wu2018aspiration, Chu2019aspiration, Zhang2019aspiration}. Win-Stay-Lose-Learn is a representative model based on aspiration, in which one will try to change its strategy only when its payoff is lower than aspiration \cite{liu2012win, chu2017win, fu2018win}. Liu and Chen investigated the Win-Stay-Lose-Learn rules in spatial prisoner¡¯s dilemma game \cite{liu2012win}; Chu and Liu added the voluntary participation into the Win-Stay-Lose-Learn rules \cite{chu2017win}; Fu studied the stochastic Win-Stay-Lose-Learn rules in the spatial public goods game \cite{fu2018win}.

These research held the assumption that an individual player¡¯s aspiration is fixed. However, according to Expected Value Theory and Achievement Motivation Theory proposed by Atkinson \cite{Atkinson1974motivation}, one¡¯s aspiration will be influenced by its previous payoff. If one¡¯s payoff is higher than its aspiration, it tends to be increased, otherwise decreased. This is consistent with our normal perception because no one can hold a high aspiration for continued low payoff.
In this paper, we thus introduce a dynamic aspiration model based on Win-Stay-Lose-Learn rules, and the principle of defection's expansion or cooperation's survival under the dynamic aspiration model is investigated.
The rest of our paper is organized as follows. First we show the detailed model for the dynamic aspiration based on Win-Stay-Lose-Learn rules. Then we give the main results under our model, which is divided into four parts: \emph{Overview}, \emph{Stable Coexistence under Low Aspiration}, \emph{Dependent Coexistence under Moderate aspiration} and \emph{Defection Explosion under High Aspiration}. Finally we discuss the wider implications of our work and the direction of the future research.

\section{Model}

Our model is described as follows.
We use the $L \times L$ square lattice with periodic boundary conditions. Each node represents a player who has one of the following two strategies: cooperation($\mathcal{C}$), or defection($\mathcal{D}$). The strategy of node $i$ is represented as $s_i$, and node $i$¡¯s aspiration is represented as $A_i$. At the beginning of the evolutionary process, each node is given an initial $s_i$ and $A_i$. The evolutionary process is performed step by step until the network is stable. In each step, players synchronously update their strategies and aspirations as follows:

(1) Each node $i$ plays the prison's dilemma game with its four neighbors and gets the payoff $P_i = \sum\limits_{j\in {\Omega_i}} P_{ij}$, where $\Omega_i$ denotes the neighbors of node $i$. $P_{ij}$ is got by Table 1, and if both of their strategies are $\mathcal{C}$ or $\mathcal{D}$, they will get the reward $R$ or punishment $P$. If $i$'s strategy is $\mathcal{C}$ and $j$'s strategy is $\mathcal{D}$, $i$ will get the suck's payoff $S$ and the $j$ will get the temptation value $T$, vice versa.

\begin{table}[h]
\caption{Payoff matrix of prison's dilemma game.}
\begin{center}
\begin{tabular}{|c|c|c|}
     \hline
     &\textbf{$\mathcal{C}$}&\textbf{$\mathcal{D}$}  \\
     \hline
     \textbf{$\quad$ $\mathcal{C}$$\quad$}&$\quad R \quad $&$\quad S \quad $ \\
     \hline
     \textbf{$\quad$ $\mathcal{D}$$\quad$}&$T$&$P$ \\
     \hline
\end{tabular}
\end{center}
\end{table}

In prison's dilemma game, the above parameters meet $\emph{T}\textgreater \emph{R}\textgreater \emph{P}\textgreater \emph{S}$.
For simplicity, we set $\emph{R}=1$, $\emph{T}=\emph{b}$, and $\emph{P}=\emph{S}=0$ in our model. This is a weak version of prisoner¡¯s dilemma
game for $\emph{P}=\emph{S}$, which is convenient but without losing the accuracy of the results \cite{nowak1992spatial}, where $1\textless \emph{b}\textless 2$ is a parameter that we will explore impact in our experiment.

(2) Each node $i$ chooses one of its four neighbors $j$ randomly with equal probability. If $i$'s payoff is lower than its aspiration, $i$ will be dissatisfied and choose to adopt $j$¡¯s strategy with the probability:

\begin{equation}
W_{ij}=\frac{1}{1+\exp[(P_i-P_j)/K]},
\end{equation}
where $K$ stands for the amplitude of noise \cite{szabo1998noise}. Without loss of generality, we use $K=0.1$ in our model \cite{szabo2005noise, perc2006noise}. If $i$'s payoff is higher than or equal to its aspiration, $i$ will be satisfied and not change its strategy.

(3) Each node $i$ updates its aspiration by the formula:

\begin{equation}
A_i(t+1) = A_i(t) + a * (P_i(t) - A_i(t)),
\end{equation}
where $A_i(0)$ is given initially. All nodes are given the same $A(0)$ and we denote it as $A$ for convenience. Based on Achievement Motivation Theory, in a representative model, one's aspiration is changed concerning its payoff linearly. So we introduce the dynamic aspiration by the parameter $a$, where $a\in[0,1]$ stands for the sensitivity of aspiration. Higher $a$ means individual's aspiration is easier to be changed by its payoff. Previous work with fixed aspiration equivalents to the case $a=0$, and $a=1$ means that individual's aspiration totally depends its payoff from the last step. Considering that one's aspiration should not be changed too drastically, we set $a=0.01$ in our model.

The step will repeat 100,000 times in one simulation. The fraction of cooperators and defectors at step $t$ are denoted as $r_{\mathcal{C}t}$ and $r_{\mathcal{D}t}$ respectively. And for each parameter, we perform 20 independent experiments to get $r_\mathcal{C}$, the fraction of cooperators when stable,  which is thought as the main index to measure a network's cooperation level.

\section{Result}

\subsection{Overview}
\begin{figure}[!htpb]

\centering

\includegraphics[width=4.5in]{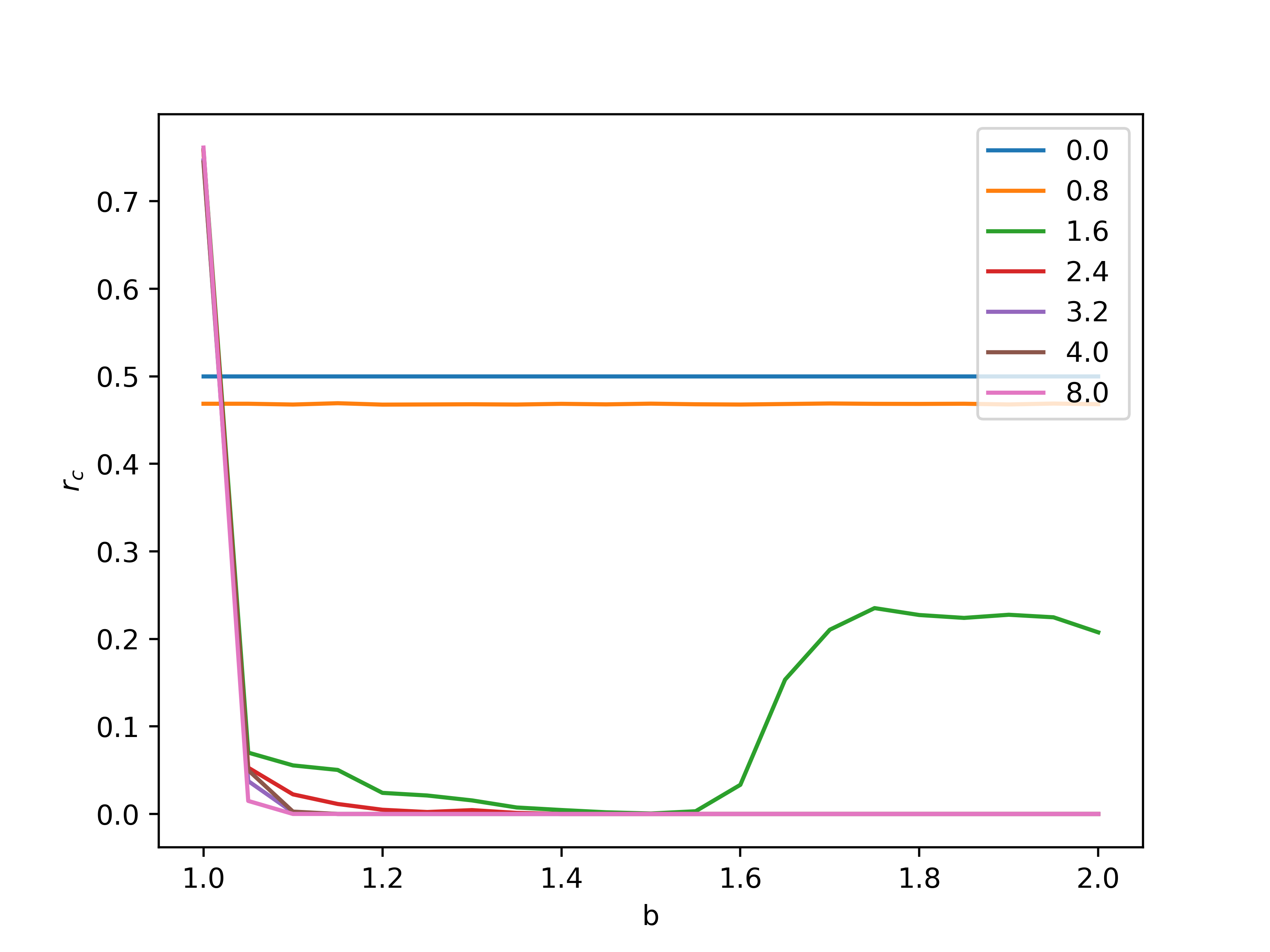}

\caption{Average fractions of cooperation when stable as a function of $b$ for different values of the initial aspiration $A$, as obtained by means of simulations on square lattices.}\label{1}

\end{figure}

Our experiment is performed on the $100 \times 100$ square lattice with periodic boundary conditions.
In initial, cooperators and defectors are distributed uniformly at random occupying half of the square lattice respectively, and all the nodes are given the same initial aspiration $A$. As the main parameters, we consider the initial aspiration level $A$ and the temptation to defect $b$. Figure \ref{1} presents the fraction $r_\mathcal{C}$ of cooperators when stable as a function of the temptation to defect $b$ for different aspiration levels. We find that the aspiration level has a significant influence on $r_\mathcal{C}$. Three different phases, \emph{Stable Coexistence under Low Aspiration}, \emph{Dependent Coexistence under Moderate aspiration} and \emph{Defection Explosion under High Aspiration} could be observed for different values of $A$.
For small values of $A$ ($A=0$ and $A=0.8$), $r_\mathcal{C}$ is equal/close to 0.5 no matter what value $b$ is.
For large values of $A$ ($A\geq2.4$), cooperators could hardly survive for any $b\textgreater1.0$.
An interesting phenomenon is discovered for moderate values of $A$ ($A=1.6$) that When $b$ is lower than 1.6, cooperators could not survive and $r_\mathcal{C}$ almost equals 0, however when $b$ is higher than 1.6, cooperation could survive and $r_\mathcal{C}$ is bigger than 0.2. According to common sense, the higher $b$ is, the harder cooperators survive, which is inconsistent with our experimental results. Above unusual phenomenon is the focus in our paper.

\subsection{Stable Coexistence under Low Aspiration ($A \leq 1.0$)}
For small values of $A$, individual's aspiration is easy to be satisfied so cooperators and defectors can coexist. For $A=0$, all the nodes are satisfied and never change their strategies, so $r_\mathcal{C}$ will always keep 0.5. And for $A=0.8$, there are only a few nodes whose four neighbors are all defectors will be dissatisfied: if it is a cooperator, it will change its strategy to $\mathcal{D}$; if it is a defector, it will be always dissatisfied but can't change its strategy since all his neighbors' strategies are the same. At the beginning, $r_{0}=r_{\mathcal{D}0}=0.5$, so the frequency of cooperators with four neighbors being all defectors could be calculated approximately:
\begin{equation}
r=r_{\mathcal{C}0}* r_{\mathcal{D}0}^4= \frac{1}{32} \approx 0.03,
\end{equation}
which is consistent with the result $r_\mathcal{C}=0.47$ in the simulation experiment and $r_\mathcal{C} \approx r_{\mathcal{C}0} - r$. In fact, the above result holds for all $A \in (0,1]$, where the proportion of cooperators dissatisfied is only $r_{\mathcal{D}0}^4$ so that cooperators and defectors could coexist. This case is called \emph{Stable Coexistence under Low Aspiration}.

\subsection{Dependent Coexistence under Moderate aspiration ($1.0 \textless A \leq 2.0$)}
\begin{figure}[!htpb]

\centering

\includegraphics[scale=0.3]{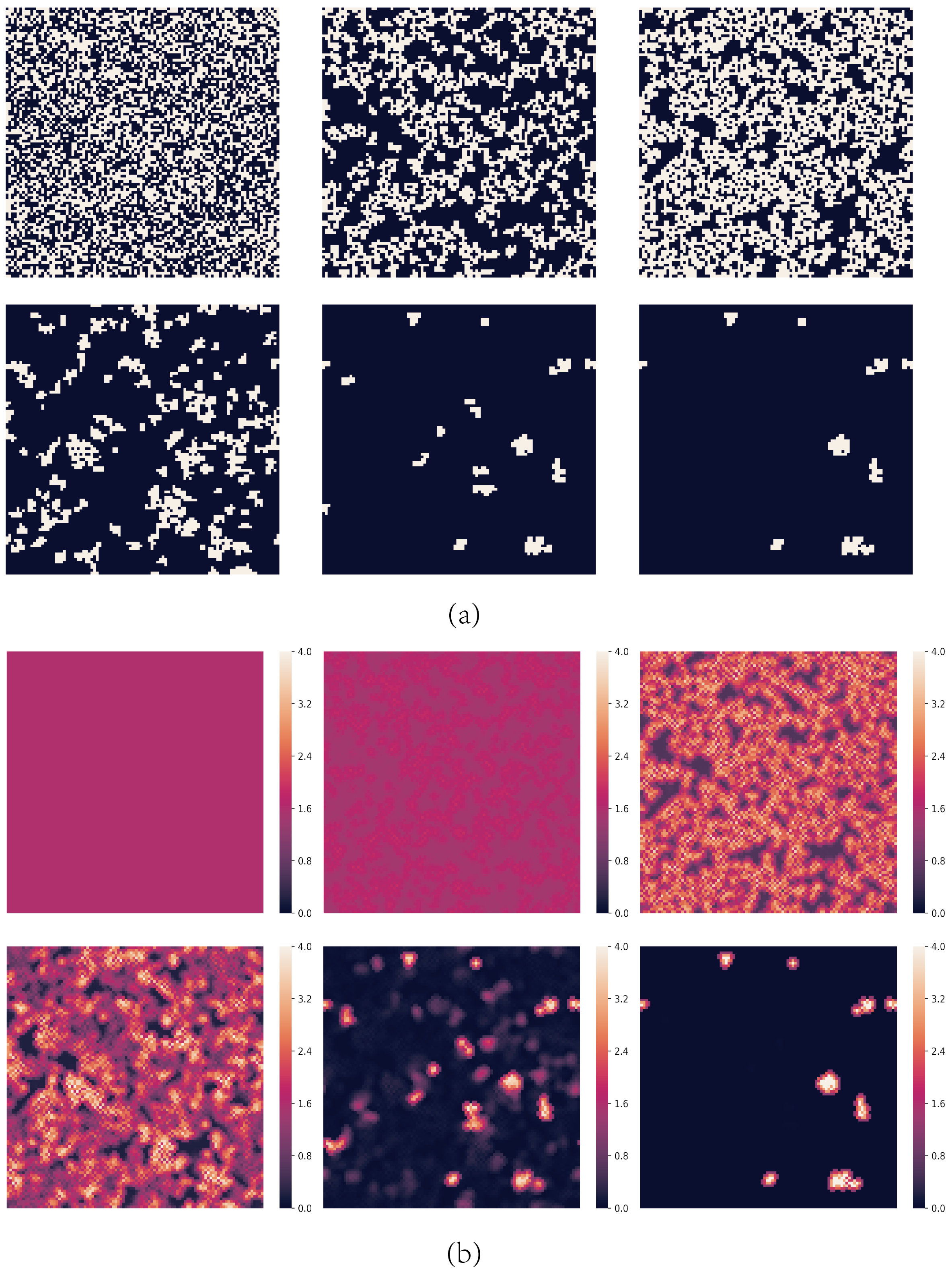}

\caption{Snapshots of typical distributions of strategies and aspirations at different time steps $t$ for $A=1.6$ and $b=1.2$. (a) represents strategies, where cooperators are depicted white and defectors are depicted black. (b) represents aspirations. The steps of them are $t$=0,10,100,200,500 and 1000 respectively.}\label{2}

\end{figure}

For moderate values of $A$, cooperators can't survive for small values of $b$.
Figure \ref{2} shows the spatial distributions of strategies and aspirations at different time steps $t$ for $A$=1.6 and $b$=1.2.
The evolution process can be divided into the following stages:

\begin{itemize}
\item
At first, every node with strictly less than two $\mathcal{C}$ neighbors is dissatisfied. Because the average payoff of defectors is higher than that of cooperators, the number of defectors increases, and the average aspiration of the network decreases.

\item
When $t=10$, cooperators that still survived have formed some small clusters so that they can survive and expand in the network, meanwhile dissatisfied defectors may evolve into cooperators gradually, and the average aspiration of the network increases.

\item
When $t=100$, cooperators occupy most of the network, however, the average aspiration of the network is higher than 2.0 and some cooperators connecting to two defectors are no longer satisfied and change into defectors. At the same time, cooperators whose neighbors are all cooperators are still satisfied so that they will not change their strategies and their aspirations will keep rising up.

\item
When $t=200$, the remaining cooperators¡¯ aspirations are close to 4.0 so that they are dissatisfied and they also change into defectors gradually, and defectors almost occupy the entire network and the cooperators almost disappear rapidly.
\end{itemize}

\begin{figure}[!htpb]

\centering

\includegraphics[scale=0.5]{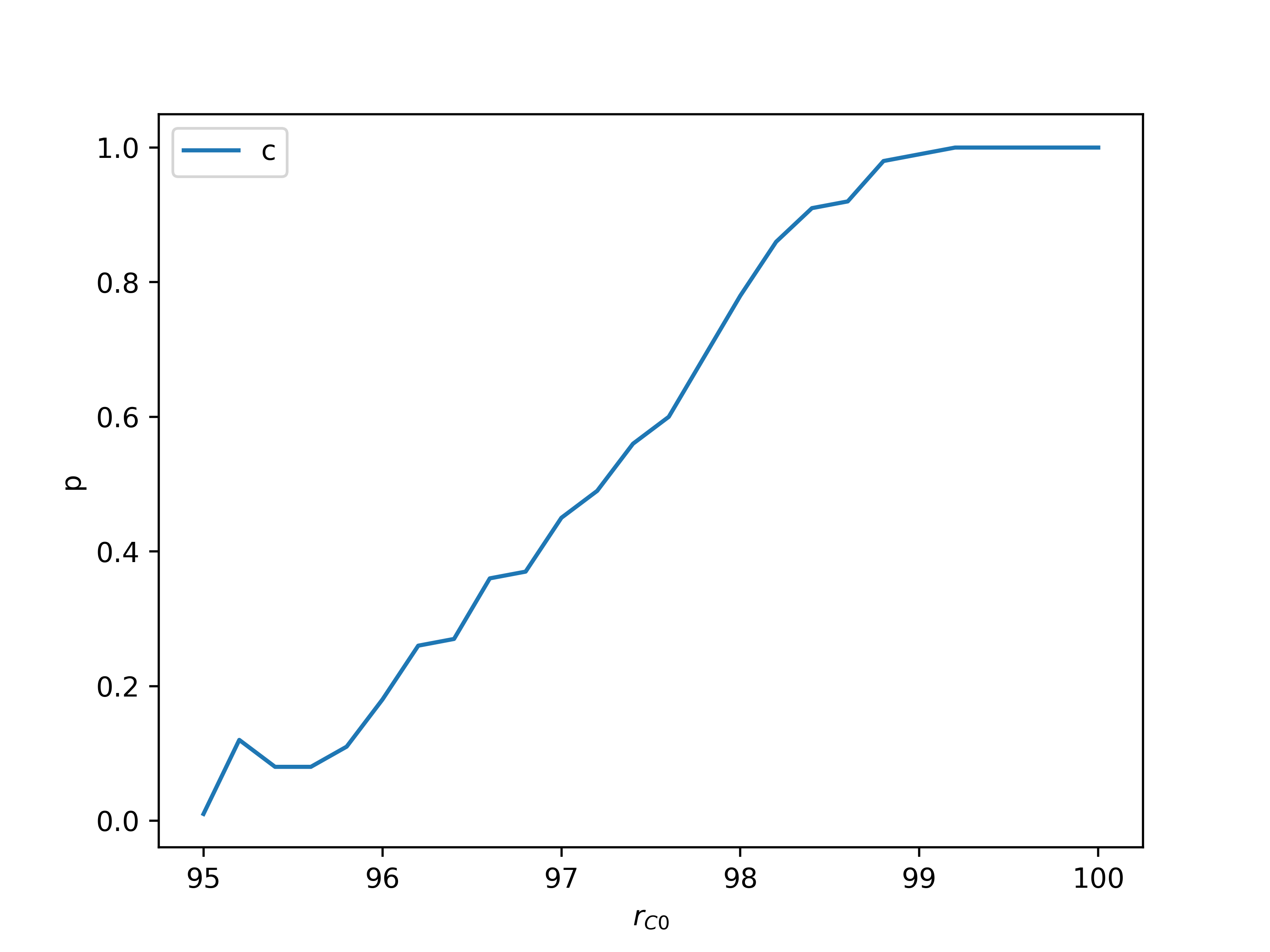}

\caption{The probability that cooperators can survive as a function of the cooperators¡¯ initial proportion $r_{\mathcal{C}0}$ for $A=1.6$ and $b=1.2$. Cooperators are easier to survive when $r_{\mathcal{C}0}$ is higher. They are possible to survive only when $r_{\mathcal{C}0} \textgreater 0.95$, and sure to survive only when $r_{\mathcal{C}0} \textgreater 0.99$.}\label{3}

\end{figure}

One can see that although cooperators take advantage in quantity in the process, defectors finally occupy the network when it is stable. As the aspirations rise up, the number of cooperators is not enough so that some cooperators are suddenly dissatisfied and evolve into defectors, which leads to the collapse of cooperation phenomenon. Figure \ref{3} shows the probability that cooperators can survive as a function of the cooperators¡¯ initial proportion $r_{\mathcal{C}0}$ for $A=1.6$ and $b=1.2$. For every different $r_{\mathcal{C}0}$, we perform 100 independent experiments. As we can see, cooperators are possible to survive only when $r_{\mathcal{C}0} \textgreater 0.95$, and sure to survive only when $r_{\mathcal{C}0} \textgreater 0.99$. In other words, defectors can survive and expand even if they are very few initially, which may be caused by some special initial structure.

\begin{figure}[!htpb]

\centering

\includegraphics[scale=0.3]{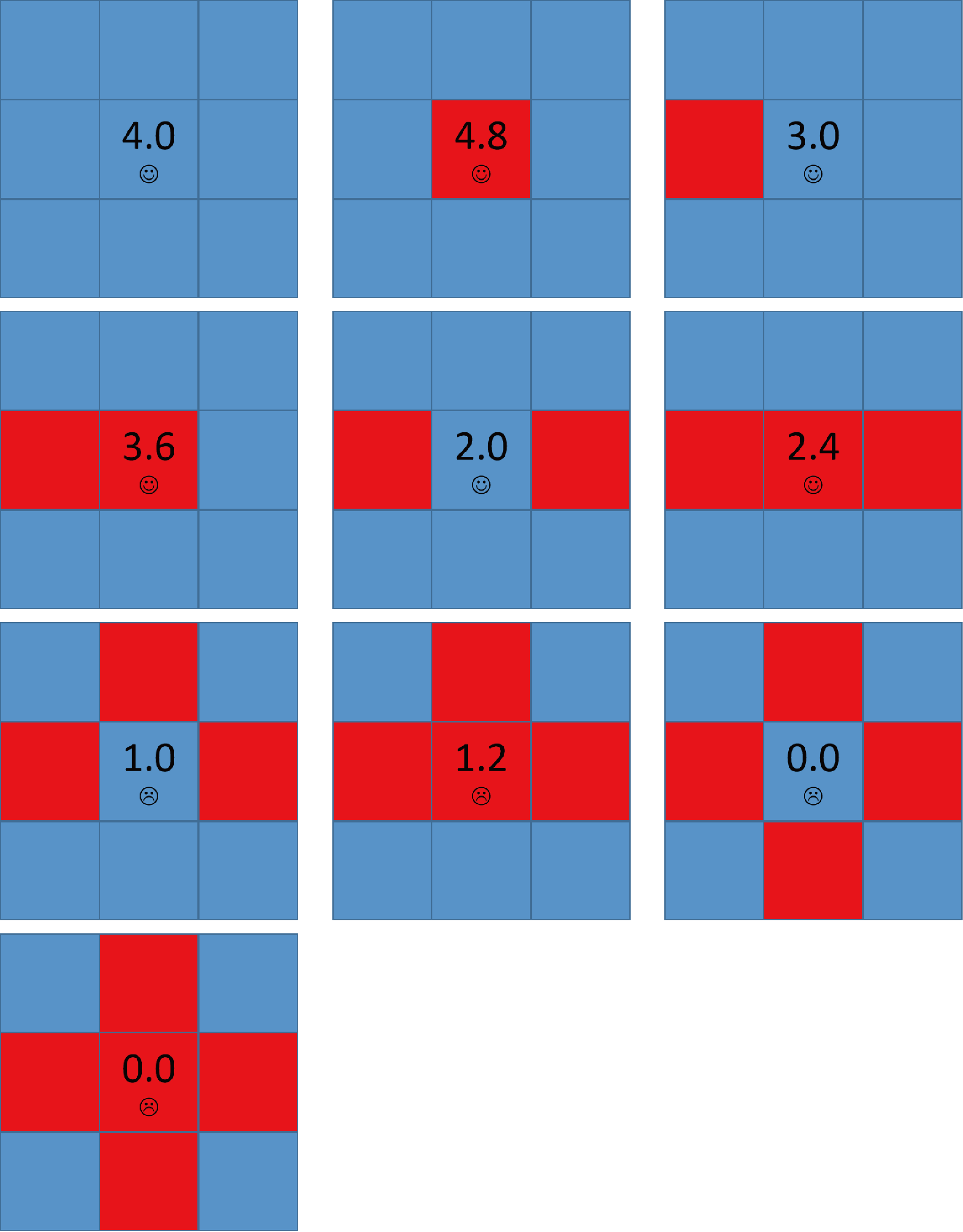}

\caption{The local structures of strategies for $A=1.6$ and $b=1.2$. Each square corresponds to a single player, where cooperators are depicted blue and defectors are depicted red. Value denoted in the center square is the individual's payoff. Smiling face represents satisfaction while crying face represents dissatisfaction.}\label{4}

\end{figure}

Figure \ref{4} shows all possible local structures in the network for $A=1.6$ and $b=1.2$. When a node has two or less $\mathcal{D}$ neighbors, it is always satisfied and won¡¯t change its strategy. When a node has four $\mathcal{D}$ neighbors, it is always dissatisfied, but since all its neighbors are defectors, it can only be a defector.

\begin{figure}[!htpb]

\centering

\includegraphics[scale=0.5]{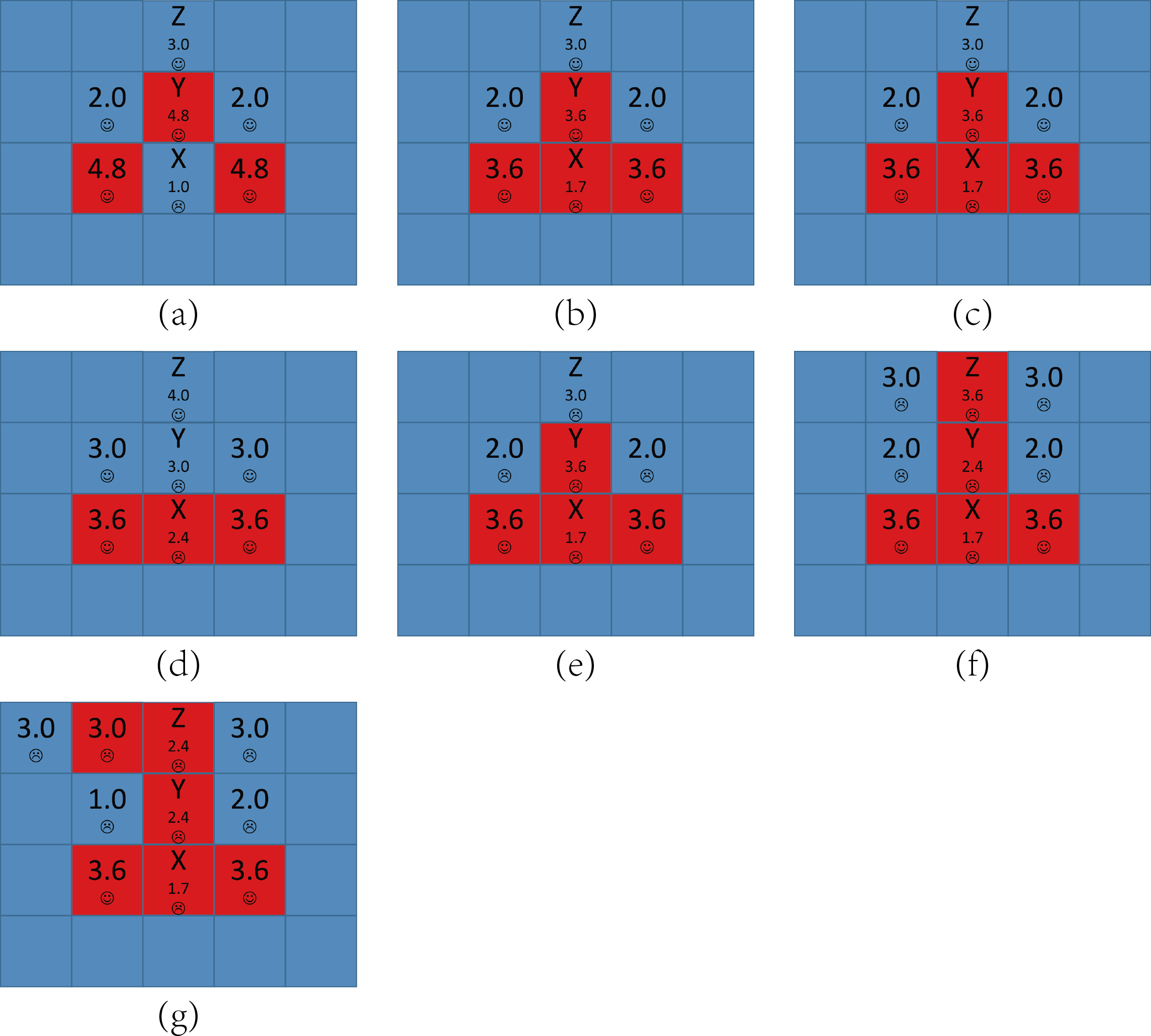}

\caption{The detailed principle for defectors' expanding for $A=1.6$ and $b=1.2$. A node is surrounded by three defectors and one cooperator initially. Smiling face represents satisfaction while crying face represents dissatisfaction.}\label{5}

\end{figure}

Now we consider the structure that a node has three $\mathcal{D}$ neighbors and one $\mathcal{C}$ neighbor. Figure \ref{5} shows the detailed principle for defectors¡¯ expanding under this initial structure.
The evolution process can be divided into the following stages:

\begin{itemize}
\item
When $t=0$, the only one node dissatisfied is node $X$ because its aspiration is 1.0. Since it has three D neighbors and one $\mathcal{C}$ neighbor, $X$ will evolve into cooperator and defector repeatedly. As a result, node $Y$¡¯s payoff is sometimes 4.8 and sometimes 3.6, and $Y$¡¯s aspiration at step $t$ can be got by the recursive equation:

\begin{equation}
A_Y(t)=\left\{
\begin{array}{lr}
A_Y(t-1)+a*(4.8- A_Y(t-1)), s_Y=\mathcal{D},   \\
A_Y(t-1)+a*(3.6- A_Y(t-1)), s_Y=\mathcal{C}
.
\end{array}
\right.
\end{equation}

\item
With $t$ growing, we can easily prove that $A_Y$ will be higher than 3.6. Next time when $X$ evolves into a defector, $P_Y=3.6\textless A_Y$, so $Y$ is dissatisfied. $Y$ has three $\mathcal{C}$ neighbors, so it will be easy to evolve into a cooperator. But when $Y$ evolves into cooperator, its payoff will decrease and it is still dissatisfied,  so $Y$ will also evolve into cooperator and defector repeatedly. For the same reason, node $Z$¡¯s payoff is sometimes 4.0 and sometimes 3.0, and $Z$¡¯s aspiration at step $t$ can be got by the recursive equation:

\begin{equation}
A_Z(t)=\left\{
\begin{array}{lr}
A_Z(t-1)+a*(4.0- A_Z(t-1)), s_Z=\mathcal{D},   \\
A_Z(t-1)+a*(3.0- A_Z(t-1)), s_Z=\mathcal{C}.
\end{array}
\right.
\end{equation}

\item
With $t$ growing, we can easily prove that $A_Z$ will be higher than 3.0. Next time when $Y$ evolves into a defector, $P_Z=3.0\textless A_Z$, so $Z$ is dissatisfied and may evolve into a defector in a few steps. Now $Z$¡¯s other neighbors¡¯ aspirations are all near to 4.0. Once $Z$ evolves into a defector, their payoffs decrease to 3.0 so they are dissatisfied and may evolve into defectors, too.

\item
Furthermore, almost every node¡¯s aspiration in the network is near to 4.0 because their payoffs have been keeping 4.0 for a long time. As a result, for each node $i$, once one of $i$¡¯s neighbors evolves into a defector, $i$ may evolve into a defector soon, which is a chain phenomenon and causes defectors¡¯ expanding.
\end{itemize}

\begin{figure}[!htpb]

\centering

\includegraphics[scale=0.3]{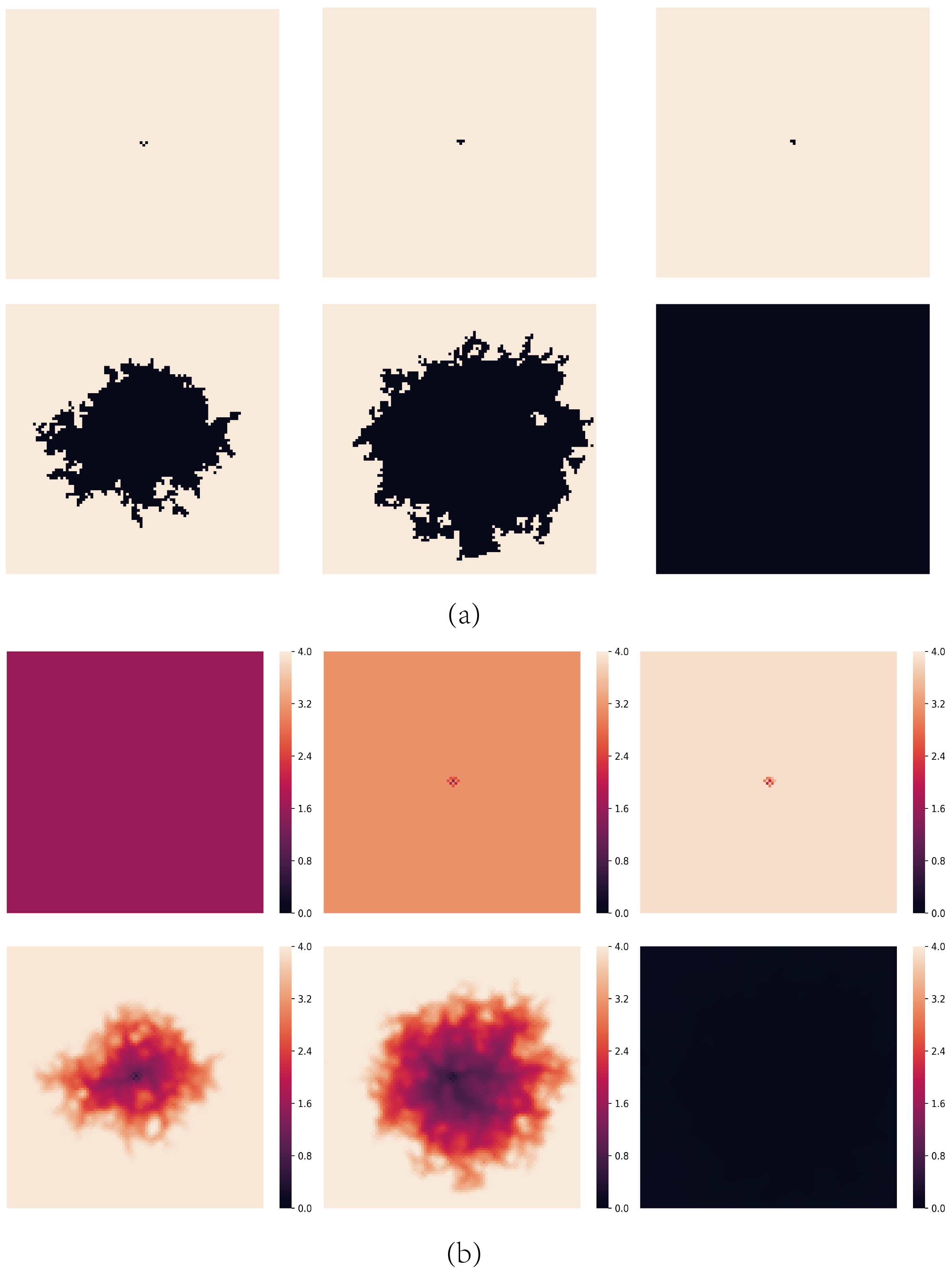}

\caption{Snapshots of typical distributions of strategies and aspirations at different time steps $t$ under the initial structure shown in Figure \ref{4} for $A=1.6$ and $b=1.2$. (a) represents strategies, where cooperators are depicted white and defectors are depicted black. (b) represents aspirations. The steps of them are $t$=0,10,100,200,500 and 1000 respectively.} \label{6}

\end{figure}

Figure \ref{6} shows the spatial distributions of strategies and aspirations at different time steps $t$ for $A=1.6$, $b=1.2$ with the above initial structure, from which we can also get the expansion trajectory by the aspiration distribution.

In the network with random setup, once there is at least one node which has three D neighbors and one C neighbor, defectors will expand to the whole network. If $r_{\mathcal{D}0} > 0.05$, such node almost certainly exists, so defectors almost certainly expand to the whole network. Cooperators can survive only when there are no above structure in the network.

\begin{figure}[htpb]

\centering

\includegraphics[scale=0.3]{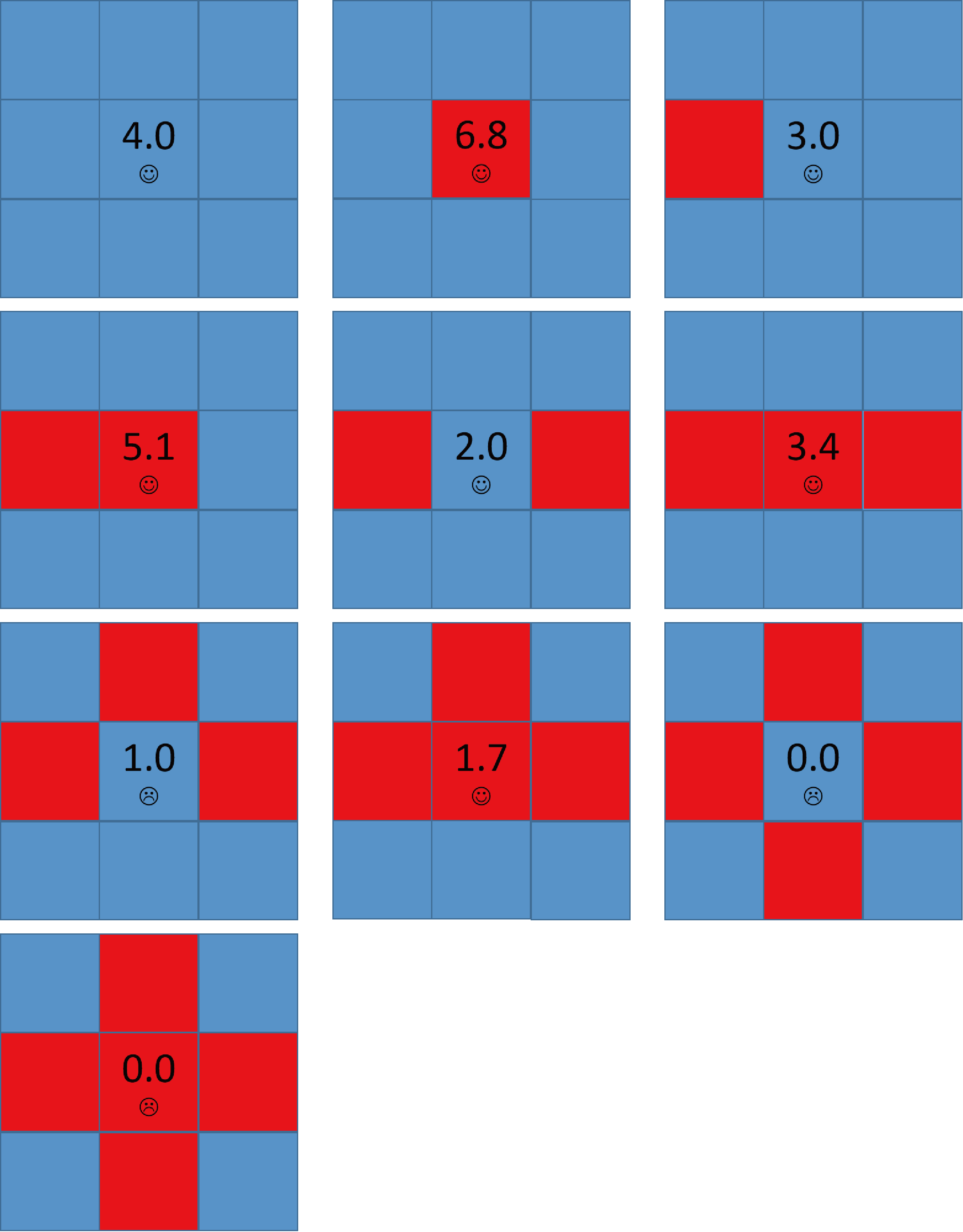}

\caption{The local structures of strategies for $A=1.6$ and $b=1.7$. Each square corresponds to a single player, where cooperators are depicted blue and defectors are depicted red. Value denoted in the center square is the individual's payoff. Smiling face represents satisfaction while crying face represents dissatisfaction.}\label{7}

\end{figure}

\begin{figure}[!htpb]

\centering

\includegraphics[scale=0.3]{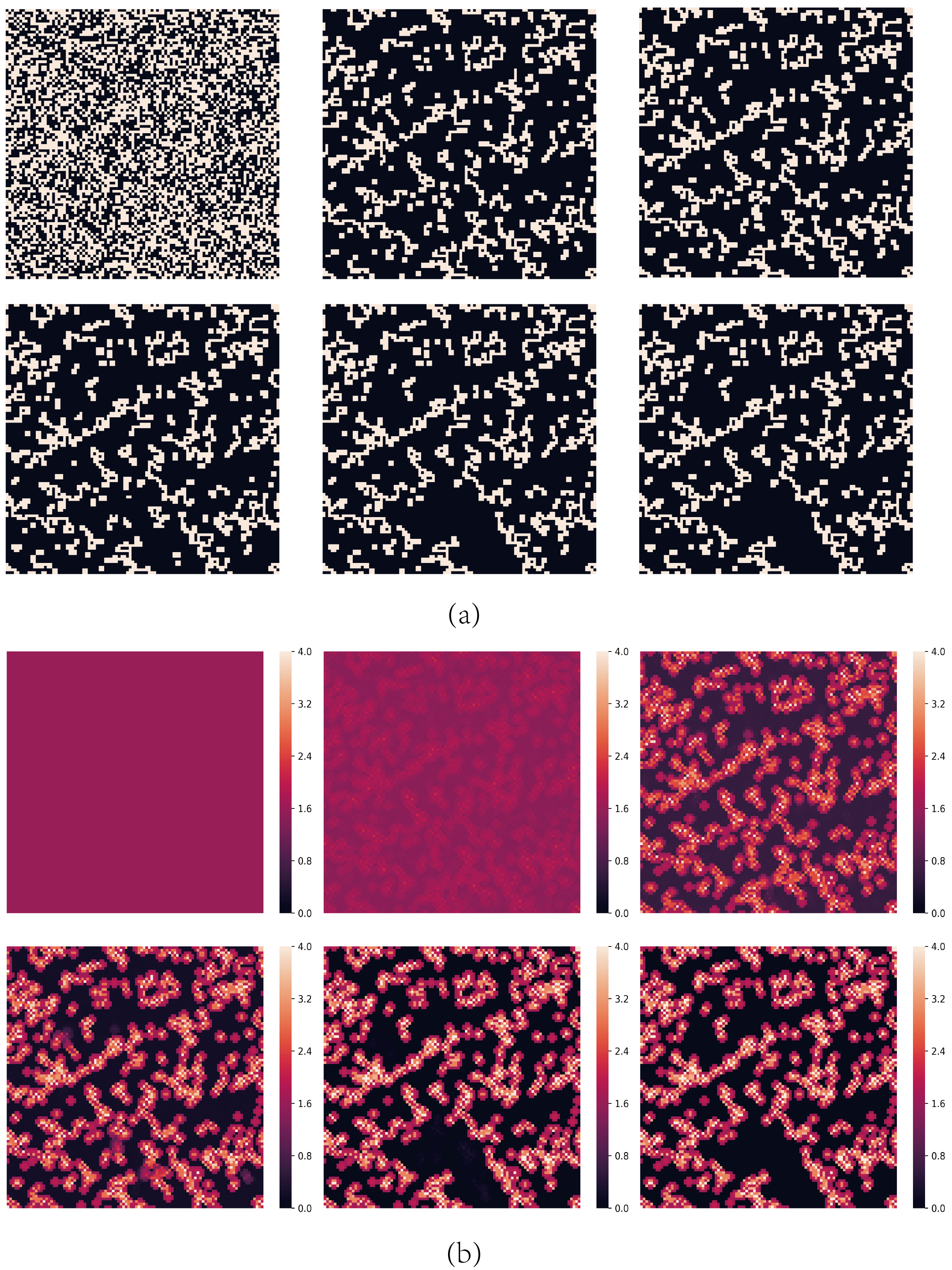}

\caption{Snapshots of typical distributions of strategies and aspirations at different time steps $t$ for $A=1.6$ and $b=1.7$. (a) represents strategies, where cooperators are depicted white and defectors are depicted black. (b) represents aspirations. The steps of them are $t$=0,10,100,200,500 and 1000 respectively.}\label{8}

\end{figure}

However for large values of $b$, cooperators can partially survive. Figure \ref{7} shows all possible local structures in the network for $A=1.6$ and $b=1.7$. Compared to Figure 4, if a cooperator has three $\mathcal{D}$ neighbors and one $\mathcal{C}$ neighbor, it will be dissatisfied and may evolve into a defector. But once it evolves into a defector, it becomes satisfied and doesn't change any longer, and the network becomes stable. Figure \ref{8} shows the spatial distributions of strategies and aspirations at different time steps $t$ for $A=1.6$, $b=1.7$ with the above initial structure. Cooperators can survive since some of them are always satisfied.

From the above we know the main difference between $b \textless 1.6$ and $b \geq 1.6$ for $A=1.6$ is whether a defector who has three $\mathcal{D}$ neighbors and one $\mathcal{C}$ neighbor is satisfied.
For $b \textless 1.6$, $P=b \textless A$, the defector is dissatisfied and will evolve into cooperator and defector repeatedly, which causes the chain phenomenon leading to defectors¡¯ expansion shown in Figure 5.
But for $b \geq 1.6$, $P=b \geq A$, the defector is satisfied and the network will be stable soon so that cooperators can survive in the end.
In fact, for all the $A \leq 2.0$, the conclusion is the same and phase transition happens in $b=A$ because when $b=A$, a defector with three $\mathcal{D}$ neighbors and one $\mathcal{C}$ neighbor has the same value of payoff and initial aspiration so that it is always satisfied.

\subsection{Defection Explosion under High Aspiration ($A \textgreater 2.0$)}

\begin{figure}[!htpb]

\centering

\includegraphics[scale=0.3]{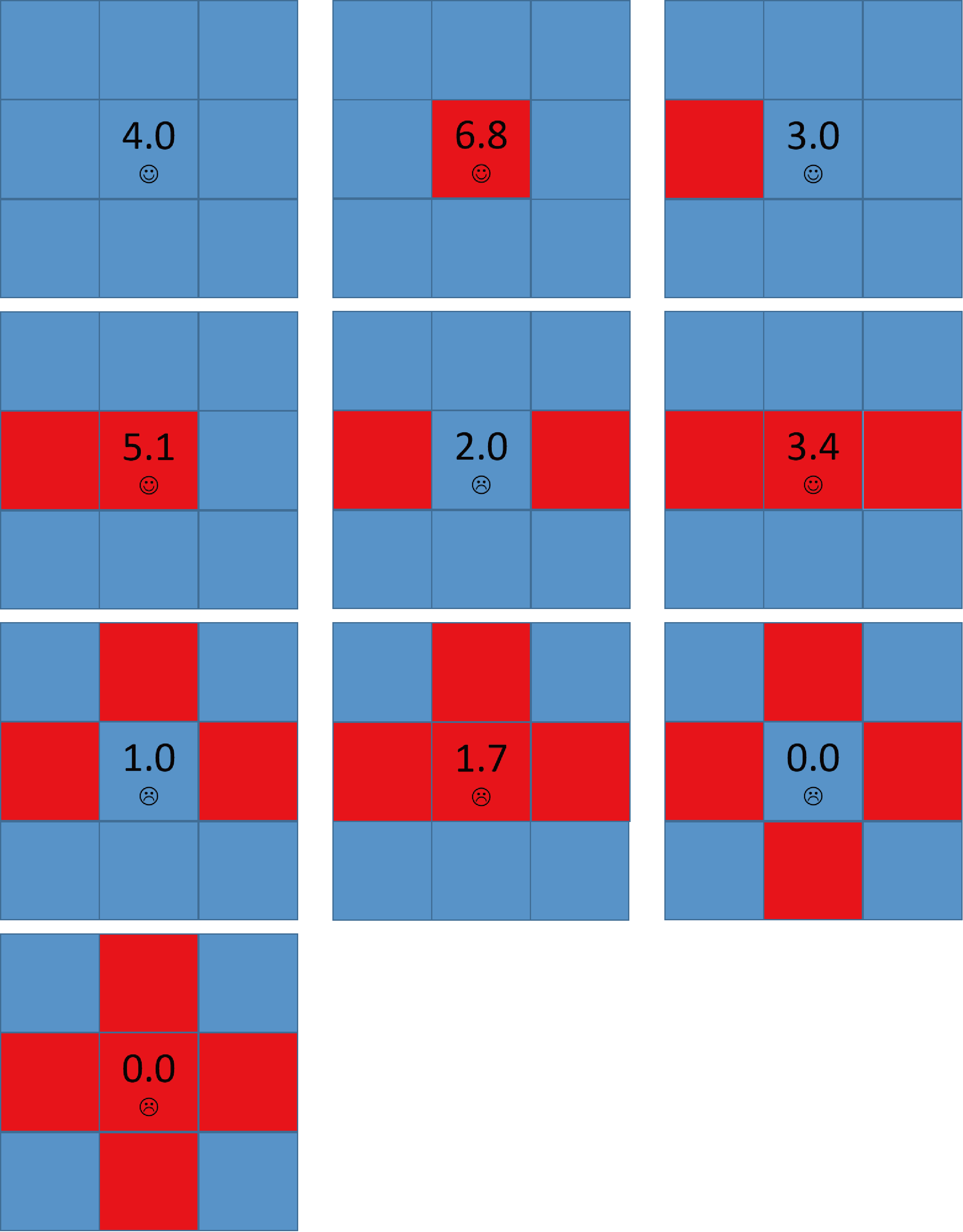}

\caption{The local structures of strategies for $A=2.4$ and $b=1.7$. Each square corresponds to a single player, where cooperators are depicted blue and defectors are depicted red. Value denoted in the center square is the individual's payoff. Smiling face represents satisfaction while crying face represents dissatisfaction.}\label{9}

\end{figure}

For $A=2.4$, cooperators can survive only when $r_{\mathcal{C}0}$ is high enough for any $b\textgreater 1$, on the contrary, only a few defectors initially can lead cooperators to die out. Figure \ref{9} shows all possible local structures in the network for $A=2.4$ and $b=1.7$. If a cooperator has two $\mathcal{D}$ neighbors, it is dissatisfied and evolves into a defector soon, and then it is satisfied. If a cooperator or defector $X$ has three $\mathcal{D}$ neighbors, it is dissatisfied and evolve into cooperator and defector repeatedly. $X$¡¯s payoff is 1.0 when it is a cooperator and 1.7 when it is a defector. $X$¡¯s aspiration at step $t$ can be got by the recursive equation:

\begin{equation}
A_X(t)=\left\{
\begin{array}{lr}
A_X(t-1)+a*(1.7- A_X(t-1)), s_Y=\mathcal{D},   \\
A_X(t-1)+a*(1.0- A_X(t-1)), s_Y=\mathcal{C}.
\end{array}
\right.
\end{equation}

With $t$ growing, we can easily prove that $A_X$ will be lower than 1.7. And next time when $X$ evolves into a defector, it will be satisfied and don't change its strategy any more. If a cooperator or defector  has four $\mathcal{D}$ neighbors, it will be never satisfied, but it can only be a defector since all its neighbors are defectors. All the simple local structures can't lead defectors to expand.

\begin{figure}[!htpb]

\centering

\includegraphics[scale=0.5]{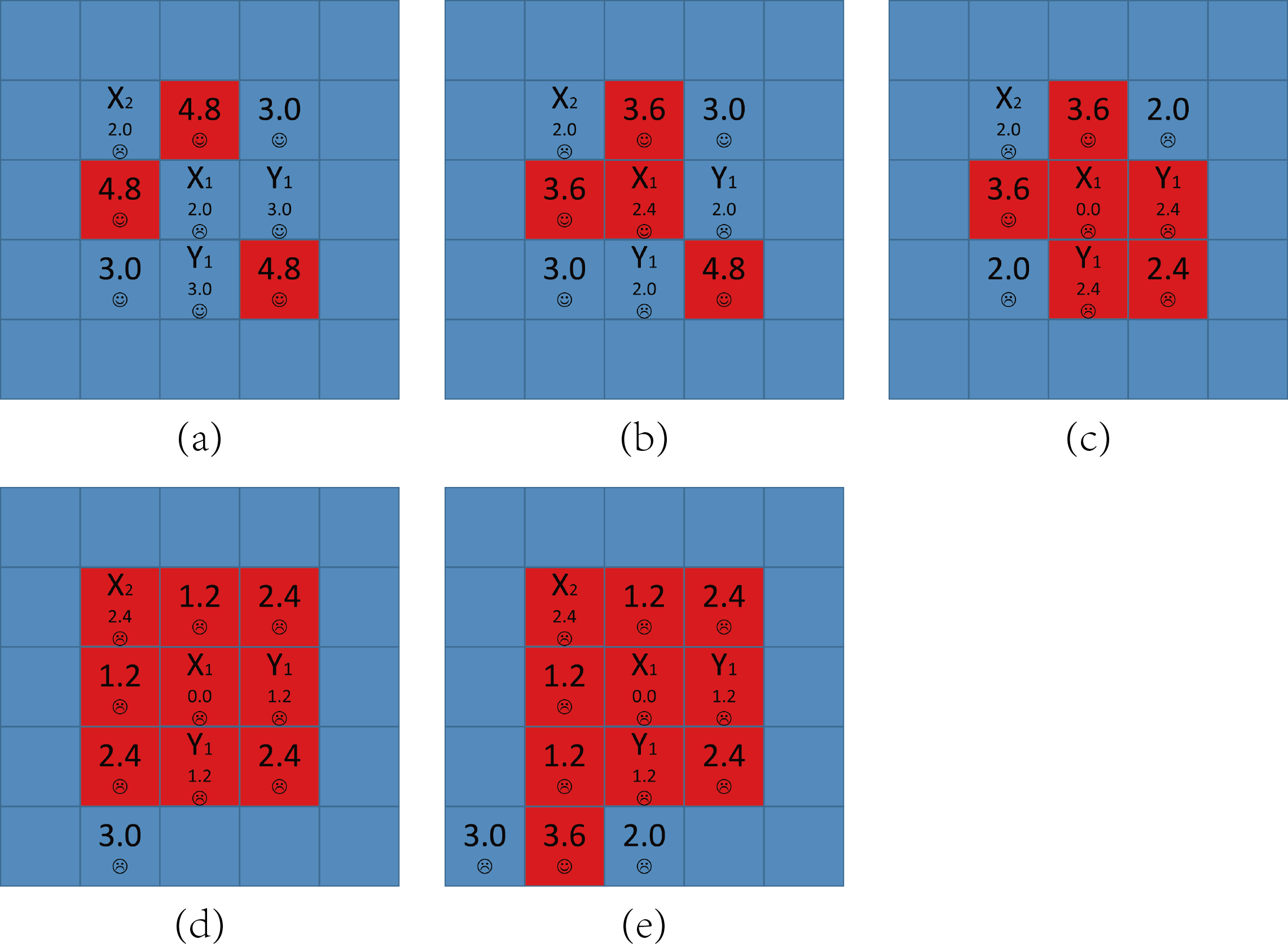}

\caption{The detailed principle for defectors' expanding for $A=2.4$, $b=1.7$. The initial local structure is shown in (a). Smiling face represents satisfaction while crying face represents dissatisfaction.}\label{10}

\end{figure}

Figure \ref{10} shows the structure which causes the defectors¡¯ expansion. In initial, nodes $X_1$ and $X_2$ are dissatisfied and may evolve into defectors. Once $X_1$ evolves into a defector, nodes $Y_1$ and $Y_2$ become dissatisfied and may also evolve into defectors, so do the other five nodes. All the nine nodes are dissatisfied and evolve into cooperators and defectors repeatedly. As a result, colored cooperators¡¯ aspiration will be higher than 3.0 as time goes. Next time when one of the nodes evolves into a defector, the cooperator will be dissatisfied and evolve into a defector. Since the other nodes¡¯ aspirations are close to 4.0, chain phenomenon occurs and defectors will occupy the whole network.

\begin{figure}[!htpb]

\centering

\includegraphics[scale=0.5]{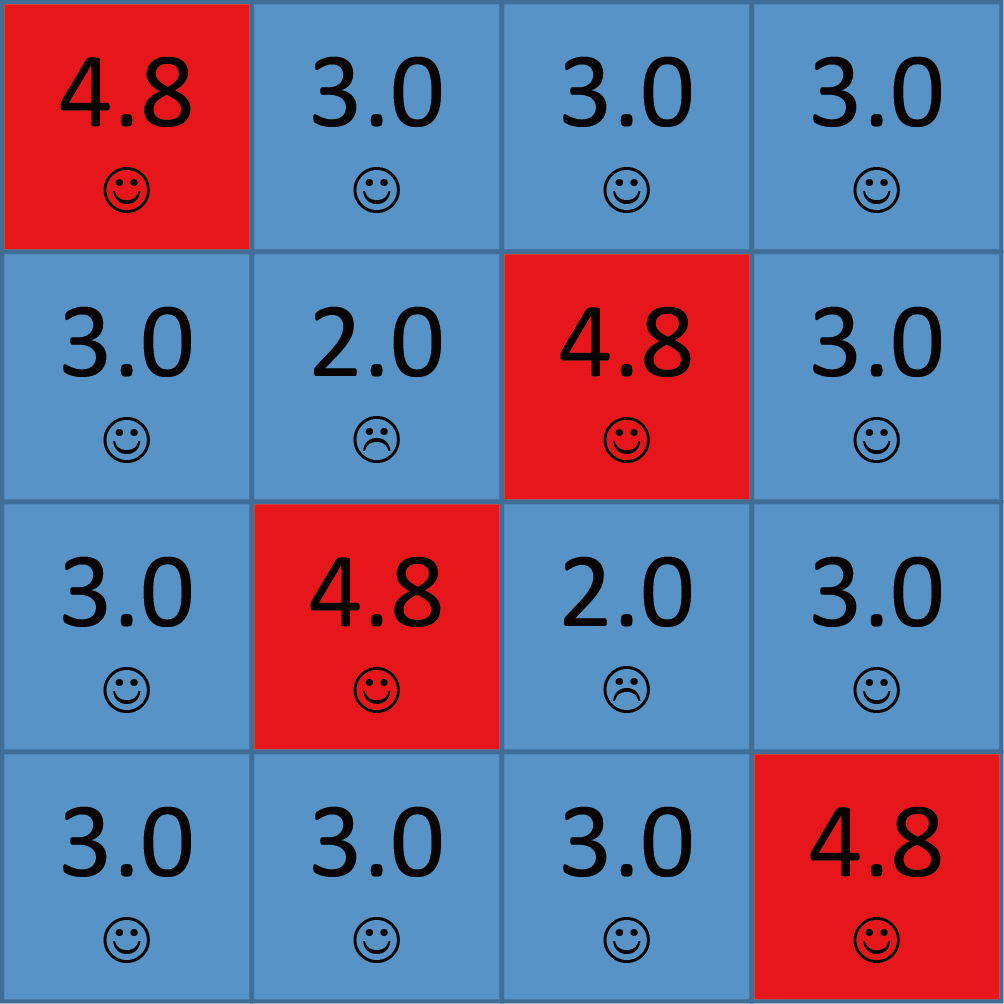}

\caption{The initial structure that causes defectors' expansion for $A=2.4$, $b=1.2$. Smiling face represents satisfaction while crying face represents dissatisfaction.}\label{11}

\end{figure}

\begin{figure}[!htpb]

\centering

\includegraphics[scale=0.3]{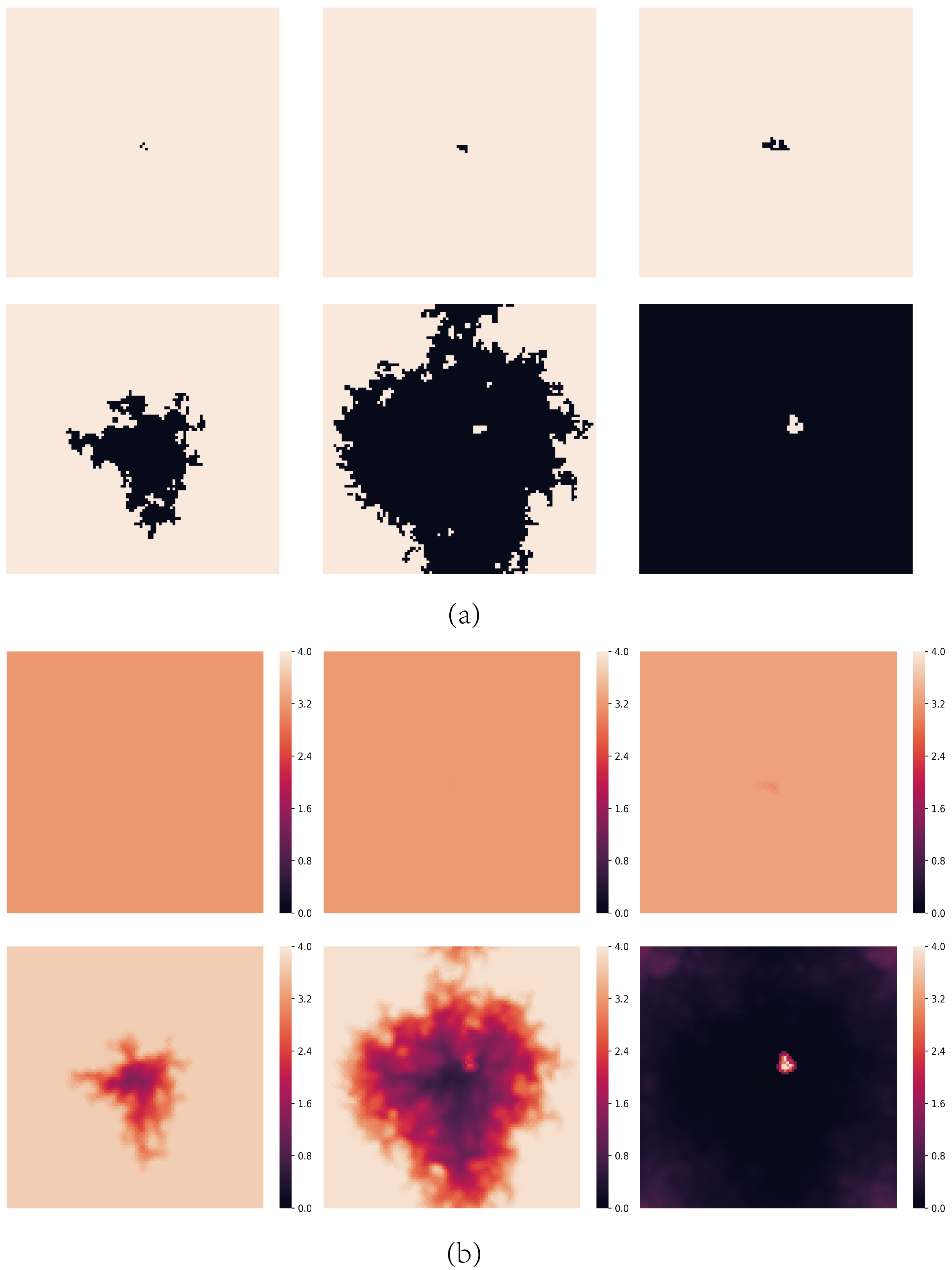}

\caption{Snapshots of typical distributions of strategies and aspirations at different time steps $t$ for $A=2.4$ and $b=1.7$ with the initial structure shown in Figure \ref{10}. (a) represents strategies, where cooperators are depicted white and defectors are depicted black. (b) represents aspirations. The steps of them are $t$=0,10,100,200,500 and 1000 respectively.}\label{12}

\end{figure}

When $b$ is lower, defectors' expansion requires more strict requirement. When $b=1.2$, for the same structure shown in Figure \ref{10}, defectors can't expand. We find that before cooperators' aspiration are higher than 3.0, the nine nodes will be all satisfied in a step so that the network becomes stable with high probability. The lower $b$ makes the network stable soon if there are only nine nodes participating in the evolution. Figure \ref{11} shows the initial structure which causes the defectors¡¯ expansion, and similar to Figure \ref{10}, sixteen nodes participate in the evolution. More active nodes make the evolutionary process lasts longer, so the colored nodes have enough time to increase their aspirations to higher than 3.0. Figure \ref{12} shows the spatial distributions of strategies and aspirations at different time steps $t$ for $A=2.4$, $b=1.7$ with the initial structure shown in Figure \ref{10}. The situation of $A=2.4$, $b=1.2$ with the initial structure shown in Figure \ref{11} is almost the same. In fact, the above conclusion is suitable for all the $2.0\textless A\leq 3.0$. Defectors' expansion requires more defectors' gathering when $b$ is lower or $A$ is higher, vice versa.

\begin{figure}[!htpb]

\centering

\includegraphics[scale=0.3]{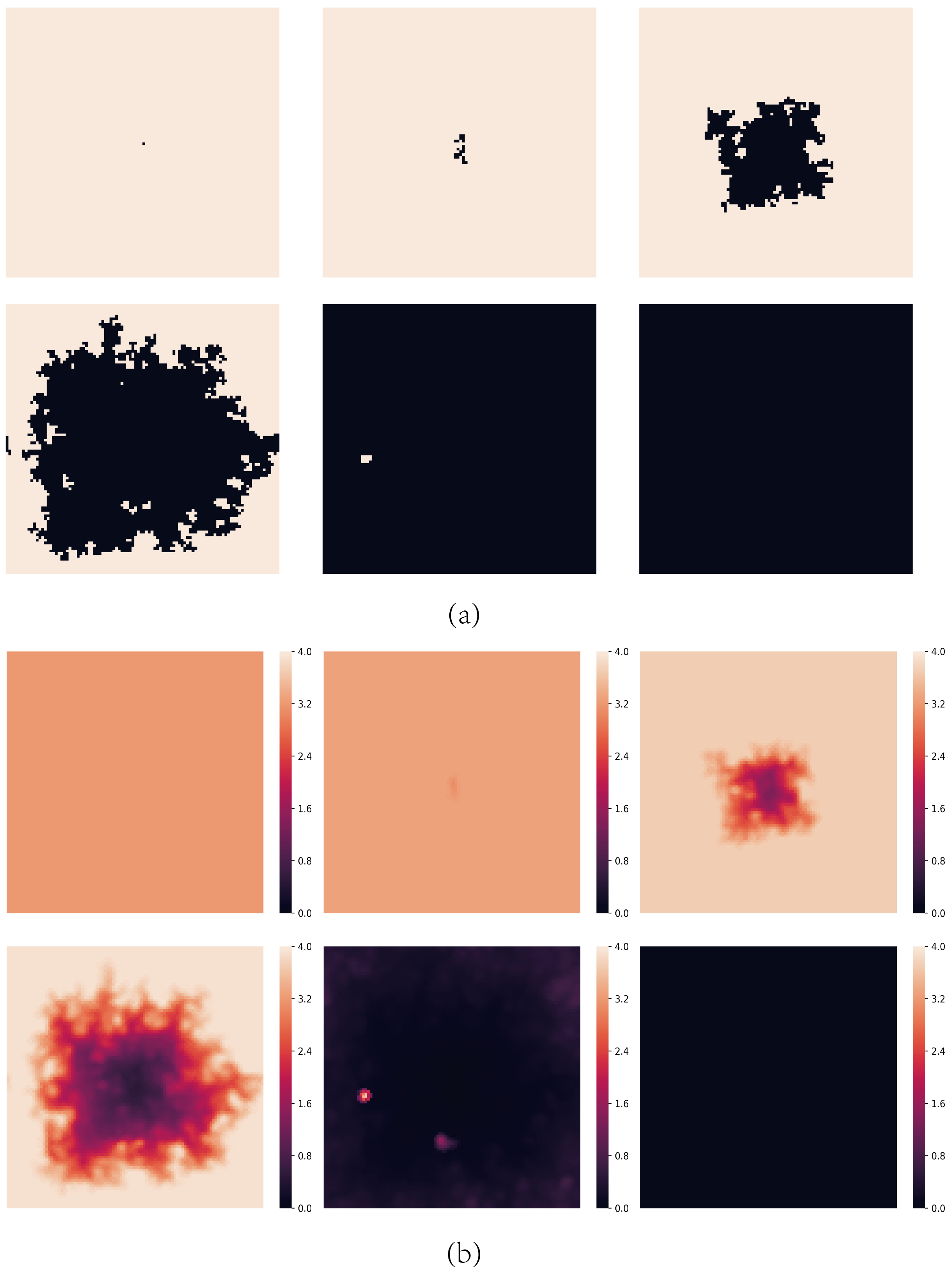}

\caption{Snapshots of typical distributions of strategies and aspirations at different time steps $t$ for $A=3.2$ and $b=1.2$ with only one defector initially. (a) represents strategies, where cooperators are depicted white and defectors are depicted black. (b) represents aspirations. The steps of them are $t$=0,10,100,200,500 and 1000 respectively.}\label{13}

\end{figure}

For $A \textgreater 3.0$, the nodes which have at least one $\mathcal{D}$ neighbor are dissatisfied, so once there are at least one defector in the network, defectors will expand to the whole network soon for any $b \textgreater 1$. The higher $b$ is, the faster defectors¡¯ expansion will be. Figure \ref{13} shows the spatial distributions of strategies and aspirations at different time steps $t$ for $A=3.2$, $b=1.2$ with only one defector initially.

In dynamic aspiration model, three different phases could be observed. The phase under low aspiration is similar to the fixed aspiration model because most nodes are always satisfied and their aspirations are changed in a small range. However, dynamic aspiration model plays a critical role under moderate aspiration and high aspiration, where some nodes are dissatisfied no matter they are cooperators or defectors and their strategies are changed repeatedly. As a result, their neighbors' payoff are changed and aspirations will be influenced by the evolution process. Their aspirations become higher gradually but their payoffs changed repeatedly, which results in their dissatisfaction. Once a node becomes dissatisfied, chain phenomenon happens and defectors will expand fast.

\section{Conclusion}
To conclude, the evolution process of the Win-Stay-Lose-Learn strategy updating rule on the prisoner¡¯s dilemma game is studied in this paper. Based on the previous work, a dynamic aspiration model is proposed, in which players will not only change their strategies based on aspirations, but also change their aspirations due to their payoffs.

Three different phases is found.
Cooperators and defectors can coexist for small values of $A$, which is called \emph{Stable Coexistence under Low Aspiration}. Only a few cooperators will evolve into defectors then and the network will be stable immediately, which is not affected by the value of $b$.
As a comparison, defectors will easily expand to the whole network for large values of $A$, which is called \emph{Defection Explosion under High Aspiration} respectively. Two kinds of local structures which can lead to defectors' expansion are found, depending on the values of $b$.
The most interesting phenomenon is cooperators can survive for higher $b$($b \geq A$) and die out for lower $b$($b \textless A$) when $1.0 \textless A \leq 2.0$, which is abnormal because higher $b$ should have meant that it is harder for the cooperators to survive, and it is called \emph{Dependent Coexistence under Moderate aspiration}. The local structure leading the defectors' expansion is that a cooperator is surrounded by one cooperator and three defectors. Dynamic aspiration plays an important role for the above results because a constantly changing individual may make its neighbors' aspirations gradually rise up and they will be satisfied no longer at a future step.


Our work provides a new enlightening opinion for the Win-Stay-Lose-Learn strategy updating rule. Dynamic aspiration introduces a more satisfactory explanation on  population evolution laws. The reason for emergence and stability of cooperative behavior under dynamic aspiration model is still a challenging problem. We hope that our work offers a valuable method that can help explore the principle behind prisoner¡¯s dilemma better, especially when combining with other rules which use aspiration level for personal decision making such as myopic, other-regarding preference or Pavlov-rule \cite{posch1999rule, fort2005rule, taylor2006rule, szabo2013rule, wu2018crule, shen2018rule}.

\section*{Acknowledgments}
This work is supported by the Research and Development Program of China£¨No.2018AAA0101100£©, the Fundamental Research Funds for the Central Universities, the International Cooperation Project No.2010DFR00700, Fundamental Research of Civil Aircraft No. MJ-F-2012-04 and the Beijing Natural Science Foundation (1192012, Z180005).

\newcommand{\newblock}{}
\bibliographystyle{unsrt}
\bibliography{ref}

\end{document}